# Anomalous magneto-thermoelectric behavior in massive Dirac materials

Yanan Li[1,10], Huichao Wang[2], Jingyue Wang[3,11], Chunming Wang[4,5], Yanzhao Liu[1], Jun Ge[1], Jingjing Niu[3], Wenjie Zhang[3], Pinyuan Wang[1], Ran Bi[3], Jinglei Zhang[6], Ji-Yan Dai[7], Jiaqiang Yan[8], David Mandrus[8,9], Nitin Samarth[10], Haizhou Lu[5,16‡], Xiaosong Wu[3,15†], Jian Wang[1,5,12,13,14*]

[1]*International Center for Quantum Materials, School of Physics, Peking University, Beijing 100871, China.*

[2]*School of Physics, Sun Yat-sen University, Guangzhou 510275, China.*

[3]*State Key Laboratory for Artificial Microstructure and Mesoscopic Physics, Peking University, Beijing 100871, China.*

[4]*Department of Physics, Shanghai Normal University, Shanghai 200234, China.*

[5]*Institute for Quantum Science and Engineering and Department of Physics, Southern University of Science and Technology, Shenzhen 518055, China.*

[6]*High Magnetic Field Laboratory, Chinese Academy of Sciences, Hefei 230031, Anhui, China.*

[7]*Department of Applied Physics, The Hong Kong Polytechnic University, Hung Hom, Kowloon, Hong Kong, China.*

[8]*Materials Science and Technology Division, Oak Ridge National Laboratory, Oak Ridge, Tennessee 37831, USA.*

[9]*Department of Materials Science and Engineering, University of Tennessee, Knoxville, Tennessee 37996, USA.*

[10]*Department of Physics, The Pennsylvania State University, University Park, PA 16802, USA.*

[11]*Center for Nanochemistry, Beijing Science and Engineering Center for Nanocarbons, Beijing National Laboratory for Molecular Sciences, College of Chemistry and Molecular Engineering, Peking University, Beijing 100871, China.*

[12]*Collaborative Innovation Center of Quantum Matter, Beijing 100871, China.*

[13]*CAS Center for Excellence in Topological Quantum Computation, University of Chinese Academy of Sciences, Beijing 100190, China.*

[14]*Beijing Academy of Quantum Information Sciences, Beijing 100193, China.*

[15]*Beijing Key Laboratory of Quantum Devices, Peking University, Beijing 100871, China.*

[16]*Shenzhen Key Laboratory of Quantum Science and Engineering, Shenzhen 518055, China.*

Extensive studies of electron transport in Dirac materials have shown positive magneto-resistance (MR) and positive magneto-thermopower (MTP) in a magnetic field perpendicular to the excitation current or thermal gradient. In contrast, measurements of electron transport often show a negative longitudinal MR and negative MTP for a magnetic field oriented along the excitation current or thermal gradient; this is attributed to the chiral anomaly in Dirac materials. Here, we report a very different magneto-thermoelectric transport behavior in the massive Dirac material $ZrTe_5$. Although thin flakes show a commonly observed positive MR in a perpendicular magnetic field, distinct from other Dirac materials, we observe a sharp negative MTP. In a parallel magnetic field, we still observe a negative longitudinal MR, however, a remarkable positive MTP is observed for the fields parallel to the thermal gradients. Our theoretical calculations suggest that this anomalous magneto-thermoelectric behavior can be attributed to the screened Coulomb scattering. This work demonstrates the significance of impurity scattering in the electron transport of topological materials and provides deep insight into the novel magneto-transport phenomena in Dirac materials.

Electrical and thermoelectric transport studies of Dirac materials are gaining increasing attention as a powerful tool to reveal the underlying physics, including the scattering mechanism, band structure, and topology. The experimental studies of Dirac materials, such as the Dirac semimetal $Cd_3As_2$ [1], the Weyl semimetal GdPtBi [2], and the massive Dirac material $Pb_{1-x}Sn_xSe$ [3], demonstrate that both the magneto-thermopower (MTP) and magneto-resistance (MR) increase with an increasing perpendicular magnetic field ($B$) [1-4], while decrease with an increasing parallel field owing to the chiral anomaly [1,2]. The classical magneto-electric conductivity under a perpendicular magnetic field is given by the Boltzmann-Drude model: $\sigma(B) = \frac{en\mu}{(1+\mu^2 B^2)}$. Here, $n$ is the carrier density; $\mu$ is the mobility with $\mu = \tau/m$ where $m$ is the effective mass and $\tau$ is the relaxation time. According to the Mott relation [5], the Seebeck coefficient can be written as $S = -\frac{\pi^2 k_B^2 T}{3e}\left(\frac{\partial \ln\sigma(E)}{\partial E}\right)_{E_F}$, where $-e$ is the electron charge, $k_B$ is the Boltzmann constant, $T$ is the temperature, $\sigma$ is the electric conductivity and $E_F$ is the Fermi level. Therefore, the detailed behavior of MTP with magnetic field (i.e. whether it increases or decreases with field) is affected by the energy dependence of the relaxation time, which depends on the impurity scattering in the system.

The fundamental understanding of impurity scattering in topological materials has been a subject of intense scientific interest. In topological insulators, impurities lead to intricate effects on both the bulk and surface properties. For example, they can modify the electrical transport [6] or induce nanoscale spatial fluctuations in the helicity of surface states [7]. In topological semimetals, the electrical conductivity can also be tuned by impurities, resulting in diverse magnetic field dependent behavior [8-13]. Recently, there has been growing interest in the thermoelectric properties of topological materials in the presence of different charge-impurity scattering mechanisms [14-19]. The thermoelectric behavior is dramatically influenced by the scattering potential and thus may reveal the nature of different scattering mechanisms in topological materials.

Amongst topological materials, the massive Dirac material $ZrTe_5$ [20-22] has been a very promising platform to explore novel quantum effects because of its small Fermi surface [23-26]. Usually, $ZrTe_5$ can reach the quantum limit at a low magnetic field, which enables the discovery of many exotic quantum phenomena, such as the log $B$ periodic quantum oscillations [27-29], the three-dimensional quantum Hall effect [30], and quantized plateau in the thermoelectric Hall conductivity [31]. Additionally, experiments have also shown anomalous transport behaviors, such as a negative longitudinal MR [32], an anomalous Hall effect [33-35] and an unconventional Hall effect [36], implying the nontrivial topological band structure of $ZrTe_5$.

In this work, we present experimental studies of combined magneto-electrical and magneto-thermoelectric transport in $ZrTe_5$ thin flakes with dominant hole carriers at low temperatures. The MR shows a magnetic field dependence that is qualitatively similar to that observed in other Dirac materials but the thermoelectric behavior shows a novel field dependence distinct from that seen in previous studies of Dirac materials. When the magnetic field is perpendicular to the thermal gradient, we observe a negative MTP that saturates in the high field regime. In addition, a positive MTP is detected when the magnetic field is parallel to the thermal gradient. Our theoretical analyses indicate that this distinct MTP behavior can be attributed to the dominant long-range screened Coulomb scattering in these crystals. Our observations are highly valuable for understanding the quantum transport phenomena in Dirac materials.

We carried out the magneto-transport measurements on $ZrTe_5$ thin flakes exfoliated from bulk crystals. Figure 1(a) shows the crystalline structure of $ZrTe_5$. Individual layers are coupled via van der Waals interactions, stacking along the $b$ axis. Our bulk $ZrTe_5$ crystals with a very low carrier density were grown by the Te-flux method as reported [37]. For transport measurements, we used electron beam lithography and electron beam evaporation to pattern electrodes on thin flakes. An excitation current $I$ or a thermal gradient $\nabla T$ was applied along the $a$ axis of $ZrTe_5$ crystals ($I//a$ or $\nabla T//a$) for electric and thermoelectric transport measurements, respectively. Figure 1(b) shows the schematic of the measurement configuration and Fig. S1(a) [38] is an optical microscopy image of a typical device. Figure 1(c) displays the $\rho$-$T$ curves from 300 K to 2 K for samples with different thicknesses. The exfoliated thin flakes clearly show a resistivity peak anomaly at a temperature that increases with decreasing flake thickness, similar to that in $ZrTe_5$ grown by the iodine vapor transfer (IVT) method [39]. We note that the peak anomaly is absent in the parent bulk crystals down to 2 K (Fig. 1(c)) while it usually presents in the IVT-grown bulk samples [40]. Additionally, the peak anomaly in the IVT-grown samples is usually accompanied by a change in the dominant carrier type [37]. However, this does not occur in our flux-grown samples. The different transport characteristics are likely related to the density of Te vacancies [37,41], and the flux-grown crystals contain relatively fewer Te vacancies (Fig. S8) [37, 38]. Figure 1(d) is the temperature dependent thermopower of a bulk crystal. The positive thermopower of both the bulk and thin flake samples (Fig. S2 [38]) indicates the dominant carriers are holes over the temperature range 300 K to 2 K, consistent with the Hall measurements (Figs. S3, S4, and S5 [38]).

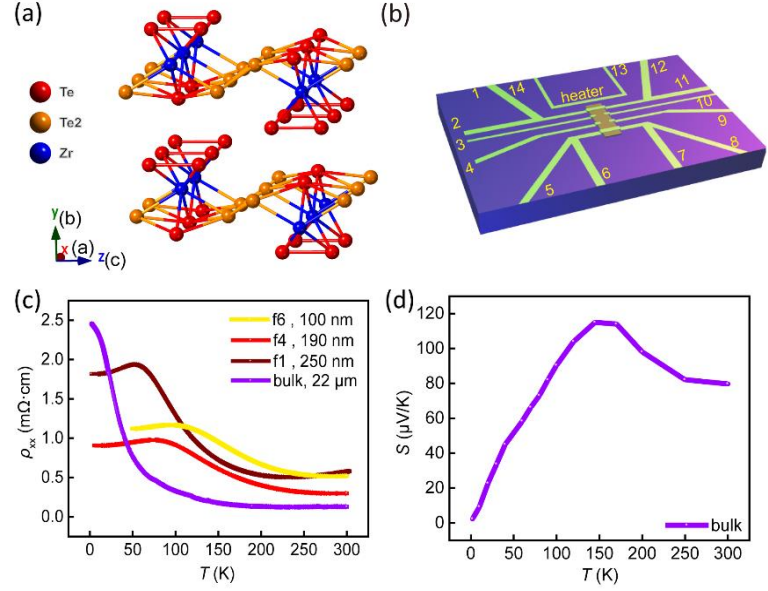

FIG. 1. Temperature dependence of the resistivity and thermopower in $ZrTe_5$. (a) Crystal structure of $ZrTe_5$. (b) Schematic of the measurement configuration. Contacts 1, 2, 11, 12 and contacts 5, 6, 7, 8 are used to inject an excitation current or measure the temperature at the two ends of the sample. Contacts 2, 3, 4, 5, 9, 10 are used to measure the MR and Hall signals in a standard Hall-bar setup. Contacts 2 and 5 or contacts 11 and 8 are used to measure the thermoelectric voltage. Contact 13 and 14 are used to apply a voltage on the heater to generate a thermal gradient on the sample. (c) The resistivity *vs*. temperature curves of the $ZrTe_5$ bulk crystal and thin flakes with different thicknesses. (d) Temperature dependence of the thermopower of a $ZrTe_5$ bulk crystal.

The magneto-transport behavior of thin flakes of $ZrTe_5$ was then measured by applying a perpendicular external magnetic field ($B//b$). In the electrical transport measurements, the samples show a positive MR with a sublinear field dependence (Figs. 2(a) and 2(b)), consistent with the previous reports [23,32,42,43]. Surprisingly, the magneto-thermoelectric transport measurements show a negative MTP in all samples. As shown in Fig. 2(c), the thermopower decreases sharply with the applied magnetic field and eventually saturates at large fields. The negative and saturated MTP extends up to 33 T (Fig. 2(d)).

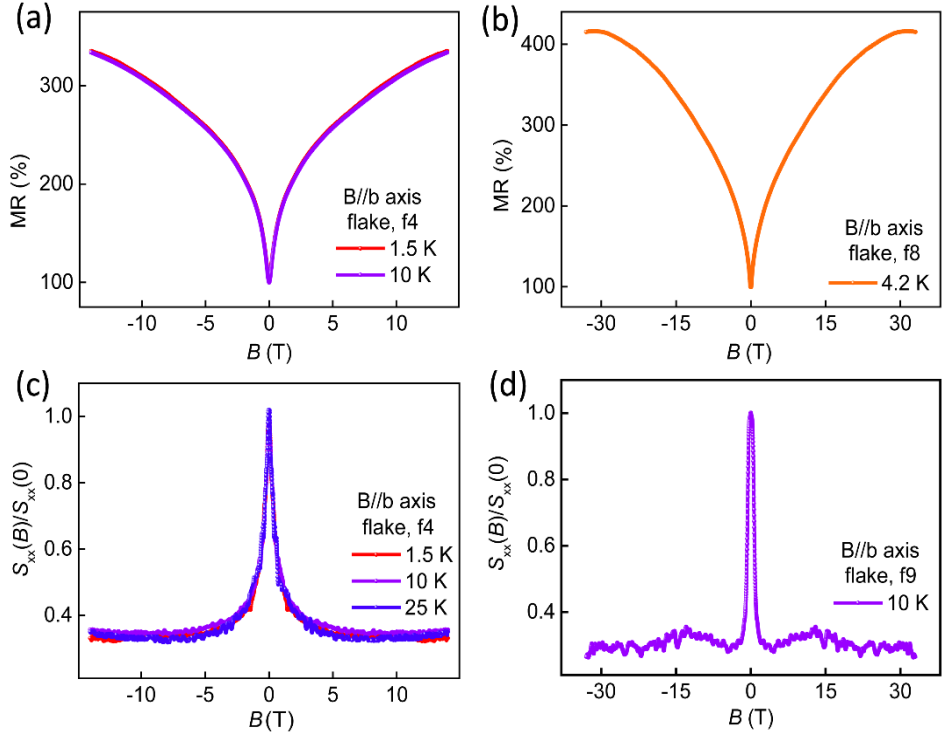

FIG. 2. Normalized MR (a-b) and MTP (c-d) in ZrTe$_5$ flake samples under perpendicular fields ($B//b$ axis). The MR is calculated as MR=R($B$)/R(0)×100%. The MTP decreases sharply at low fields and saturates at high magnetic fields.

As the magnetic field orientation is tilted from the perpendicular configuration ($B//b$ axis) to the current direction ($B//a$ axis), the positive MR is first suppressed and then flips to the negative MR, as shown in Figs. 3(a) and 3(b). This behavior is further confirmed in another sample (Fig. S6(a) [38]). The amplitude of the negative MR generally decreases with increasing temperature and vanishes at a temperature between 100 K and 150 K, as shown in Fig. 3(b). The negative MTP is also suppressed when the field orientation deviates from $b$ axis, and a positive MTP is observed when the magnetic field is tilted to the temperature gradient direction (Figs. 3(c) and 3(d)). The flat trace at low fields when $\theta=90°$ (Fig. 3(c)) may be due to the presence of a tiny misalignment angle in the measurement, i.e., the negative MTP contribution from the perpendicular component offsets the positive MTP under the parallel field [38]). In another sample (Fig. S7 of Supplemental Material [38]), we observe clear positive MTP under parallel fields in the whole field range. The amplitude of the positive MTP also decreases gradually with increasing temperature and survives up to 100 K (Fig. 3(d)). The longitudinal magneto-thermoelectric conductance of the material is calculated based on the formula $\alpha(B) = S(B)\sigma(B)$ and shown in Figs. 3(e, f). Here, $\alpha(B)$ denotes the magneto-thermoelectric conductance. Thus, in addition to the positive longitudinal magneto-conductance, we also unveil a positive longitudinal magneto-thermoelectric conductance with a similar temperature and angle dependence as the positive MTP.

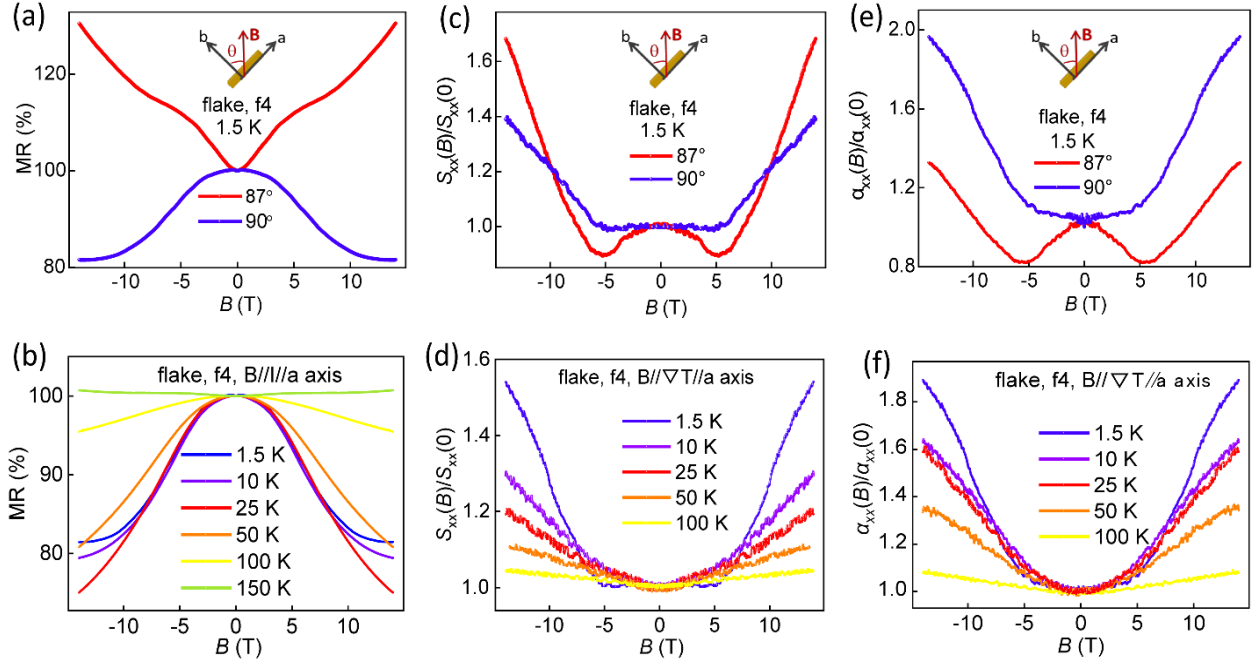

FIG. 3. Negative MR and positive MTP in ZrTe$_5$ under parallel magnetic fields ($B//I$ or $B//\nabla T$). (a, c) Angle dependence of normalized MR and MTP of ZrTe$_5$ thin flakes near the parallel magnetic fields. (b, d) Temperature dependence of the normalized MR and MTP of ZrTe$_5$ thin flakes under parallel fields. (e, f) Angle dependence and temperature dependence of the magneto-thermoelectric conductance calculated by the formula $\alpha(B) = S(B)\sigma(B)$.

Figures 2 and 3 show that we have detected rather anomalous magneto-thermoelectric behavior in the flux-grown ZrTe$_5$. Recent studies of bulk crystals and flakes (a few hundred nanometers thick) have shown that ZrTe$_5$ is a massive Dirac material which holds three-dimensional massive Dirac fermions with nearly linear bulk band dispersion and a bandgap [21,44]. To interpret the results, we use the massive Dirac model which has been effectively used to describe ZrTe$_5$ in previous studies [21]. Previous magneto-infrared spectroscopy measurements suggest that the quantum limit of flux-grown ZrTe$_5$ bulk crystals and thin flakes is lower than 1 T [20,45]. In our measurements, the maximum magnetic field is 33 T and thus we focus on the quantum transport and scattering mechanism beyond the quantum limit. In perpendicular magnetic fields, the longitudinal conductivity $\sigma_{xx}$ from the zeroth Landau level is related to the relaxation times $\tau_{1s\lambda}$, which arises from the virtual scattering processes between the zeroth Landau band $E_{0+}$ and the first Landau bands $E_{1s\lambda}$. $\sigma_{xx}$ is then given by [46,47]

$$\sigma_{xx} \simeq \frac{\hbar e^2}{2\pi} \frac{1}{2\ell_B^2} \int_{-\infty}^{\infty} d\varepsilon \left[ -\frac{\partial n_F(\varepsilon)}{\partial \varepsilon} \right] \sum_{s,\lambda,k_z} \left[ (v_{0,1s\lambda}^x)^2 \frac{\frac{\hbar}{2\tau_{1s\lambda}}}{(\varepsilon - E_{1s\lambda})^2 + \left(\frac{\hbar}{2\tau_{1s\lambda}}\right)^2} \delta(\varepsilon - E_{0+}) \right]. \quad (1)$$

Here $\ell_B = \sqrt{\hbar/eB}$ is the magnetic length, $n_F(\varepsilon)$ is the Fermi distribution function. The velocity element $v_{0,1s\lambda}^x$ couples the zeroth band and bands $1s\lambda$. In perpendicular magnetic fields, we also need to consider the Hall conductivity. Then, the resistivity and the Seebeck coefficient are given by $\rho_{xx} = \sigma_{xx}/(\sigma_{xx}^2 + \sigma_{xy}^2)$ and

$$S_{xx} = \frac{\pi^2 k_B^2 T}{3e} \frac{1}{\sigma_{xx}^2 + \sigma_{xy}^2} \left( \sigma_{xx} \frac{\partial \sigma_{xx}}{\partial E_F} + \sigma_{xy} \frac{\partial \sigma_{xy}}{\partial E_F} \right). \quad (2)$$

The relaxation time $\tau_{1s\lambda}$ depends on the electron-impurity scattering and can be calculated in the Born approximation [46,47] as

$$\frac{\hbar}{\tau_{1s\lambda}} = \pi n_i \ell_B^2 \sum_{k_z'} \left( F_{k_z,k_z'}^{1s\lambda,0} \right)^2 \sum_q |u(q)|^2 q_\perp^2 e^{-\ell_B^2 q_\perp^2/2} \delta_{q_z,k_z - k_z'} \delta\left(\varepsilon - E_{0+}^{k_z'}\right) \quad (3)$$

Here, $n_i$ is the impurity density, $F_{k_z,k_z'}^{1s\lambda,0}$ is a form factor, $u(q)$ is the Fourier transformation of the scattering potential, $q_\perp^2 = q_x^2 + q_y^2$, and $E_{0+}^{k_z'} = \sqrt{v^2(k_z')^2 + (b_z - m)^2}$ is the energy dispersion of the lowest Landau band with $v$ the Fermi velocity, $b_z$ the Zeeman splitting energy, and $m$ the gap.

It is known that the impurity scattering potential could strongly influence the thermoelectric transport behavior. We have examined two typical scattering potentials, the screened Coulomb potential induced by charged impurities and the Gaussian scattering potential induced by neutral impurities. Since the anomalous thermoelectric behavior is stronger at low temperature, scattering from phonons is excluded from our analyses. The screened Coulomb potential of charged impurities is $U(r) \propto \sum_i e^{-\kappa|r-R_i|}/|r - R_i|$, while the random Gaussian scattering potential from the neutral impurities is $U(r) \propto \sum_i e^{-|r-R_i|^2/2d^2}$. Here, $R_i$ is the position of a randomly distributed impurity, $1/\kappa$ is the screening length for screened Coulomb potential, and $d$ is the acting range of the impurity potential for Gaussian potential. Our calculations (Figs. 4(a) and (b)) reveal that both scatterings yield positive MR and negative MTP as observed in the experiments when the potentials are long-ranged, that is, $1/\kappa, d > \ell_B$ (see Pt. IV of Supplemental Material for more details [38]).

In a parallel field, the conductivity is related to the diagonal element of the velocity and the vertex correction of the velocity element should be considered. After vertex correction, the relaxation time is corrected to the transport time [48,49], giving the following formula

$$\sigma_{xx} = \frac{e^2}{h} \frac{1}{2\pi \ell_B^2} \frac{v^2 k_F}{E_F} \left( \frac{\tau_{k_F}^{tr}}{\hbar} + \frac{\tau_{-k_F}^{tr}}{\hbar} \right). \quad (4)$$

The transport time at the Fermi energy $E_F$ is given by

$$\frac{\hbar}{\tau^{tr}_{\pm k_F}} = \frac{2E_F}{v^2 k_F} n_i \cos^2\left[\frac{1}{2}(\alpha_{k_F} - \alpha_{-k_F})\right] \sum_{q_y,q_z} |u(\pm 2k_F, q_y, q_z)|^2 e^{-(q_y^2+q_z^2)\ell_B^2/2}. \quad (5)$$

with $\alpha_{\pm k_F} = \mp tan^{-1}[vk_F/(b_x + m)]$, $k_F$ the Fermi wave vector, $b_x$ the Zeeman energy. The resistivity is the inverse of the conductivity and the Seebeck response can be obtained through the Mott relation. The transport time strongly relies on the scattering potential, which results in various magnetic-field dependences for various scattering types. In particular, the resistivity due to the Gaussian potential at fixed carrier density is given by:

$$\rho_{xx} = \frac{h}{e^2} \frac{n_i u_0^2}{v^2} \frac{(b_x+m)^2}{4\pi^4 v^2 n^2 \ell_B^2 (2d^2+\ell_B^2)} e^{-16\pi^4 n^2 \ell_B^4 d^2}. \quad (6)$$

Since $b_x$ increases with the magnetic field, while $l_B$ decreases, $\rho_{xx}$ always increases with the field for arbitrary acting ranges (Fig. 4(d)), conflicting with the experimental results. However, the screened Coulomb scattering potential can lead to positive MTP and negative MR regardless of screening length in parallel fields (Fig. 4(c) and Fig. S10 [38]), consistent with our observations. This parallel field case is completely distinct from the perpendicular one. In perpendicular magnetic field, the in-plane motion is quantized and the in-plane momentum is not a good quantum number. The velocity is off-diagonal. The conductivity relates to the higher-order off-diagonal velocity element, which leads to the virtual process going back and forth between the zeroth band and the nearest high band [46]. The conductivity is almost inversely proportional to the relaxation time. Further, the Hall conductivity also influences the magnetoresistance and the Seebeck coefficient in perpendicular field case. Combining the calculations and experimental results, we qualitatively attribute the observed anomalous MTP and MR phenomena to the transport behavior in the presence of long-range screened Coulomb potential. More details of the calculations are shown in the Supplemental Material [38].

In the quantum limit, all the electrons occupy the lowest Landau band. This band is affected by the external magnetic field. In contrast to the classical case, the transport time or the relaxation time of this one-band quantum system is strongly influenced by the magnetic field and the scattering potential. Therefore, the transport behavior in the quantum limit is very sensitive to the scattering mechanism. As shown by the calculations, impurity scattering strongly influences the relaxation time (perpendicular field configuration) or transport time (parallel field configuration) in massive Dirac materials, which determines the magneto-transport properties. In the ZrTe$_5$ samples grown by the IVT method, the MTP shows a nonmonotonic field dependence in perpendicular field and a negative field dependence in parallel field [50,51]. The difference in thermoelectric transport behavior between IVT-grown ZrTe$_5$ and flux-grown ZrTe$_5$ may originate from the different impurity scattering potentials formed during the growth process. Furthermore, our results suggest a need for caution in interpreting a negative MR under a parallel magnetic field as the signature of a chiral anomaly in Dirac materials [1] or helicity transport in massive Dirac materials [52]. Our studies indicate that impurity scattering should also be taken into serious consideration in the understanding of the anomalous transport behaviors.

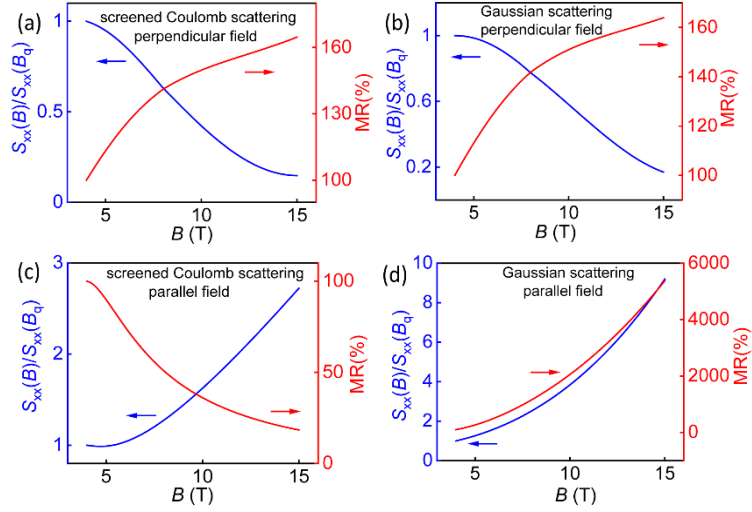

FIG. 4. MR and MTP calculated using the massive Dirac model and different scattering potentials. The MR and MTP are normalized with respect to the value at $B_q$ (quantum limit). (a, c) The results calculated using the long-range screened Coulomb scattering. A negative MTP and a positive MR are obtained under the perpendicular field, while a positive MTP and a negative MR are obtained under the parallel field, consistent with the experimental observations. (b, d) The results calculated using the Gaussian scattering. The MR is positive for the parallel field case, contrary to the experimental observations.

In conclusion, we observed anomalous magneto-thermoelectric transport behavior in thin flakes of massive Dirac material $ZrTe_5$. We found a large negative MTP in a perpendicular field while the MTP was positive in a parallel field. The MTP behavior is distinct from the generally expected results in Dirac materials although the observed MR shows a field dependence consistent with previous reports. This exotic behavior can be qualitatively explained by the long-range screened Coulomb scattering potential in the quantum limit of a massive Dirac band structure. The combined angle-dependent electrical and thermoelectric transport studies presented in this work provide important new insights into quantum transport phenomena in topological materials.

We thank Xiaobin Qiang and Chuanying Xi for helpful discussions. J.W. acknowledges the National Key R&D Program of China (2018YFA0305604), Beijing Natural Science Foundation (Z180010), the National Natural Science

Foundation of China (11888101), the Strategic Priority Research Program of Chinese Academy of Sciences (XDB28000000). X.W. acknowledges the support from the National Key Basic Research R&D Program of China (2020YFA0308800) and the National Natural Science Foundation of China (11774009, 12074009). H.L. acknowledges the National Natural Science Foundation of China (11925402), Guangdong province (2016ZT06D348, 2020KCXTD001), Shenzhen High-level Special Fund (G02206304, G02206404), and the Science, Technology and Innovation Commission of Shenzhen Municipality (ZDSYS20170303165926217, JCYJ20170412152620376, KYTDPT20181011104202253), and Center for Computational Science and Engineering of SUSTech. H.W. acknowledges the National Natural Science Foundation of China (12004441, 92165204), the Hundreds of Talents program of Sun Yat-sen University and the Fundamental Research Funds for the Central Universities (202lqntd27). C.W. acknowledges the National Natural Science Foundation of China (11974249) and the Natural Science Foundation of Shanghai (19ZR1437300). J.Y. and D.M. acknowledge the U.S. Department of Energy, Office of Science, Basic Energy Sciences, Materials Sciences and Engineering Division.

Y.L., H.W., Jingyue W and C.W. contributed equally to this work.

*Corresponding author.

  jianwangphysics@pku.edu.cn (J.W.)

†Corresponding author.

  xswu@pku.edu.cn (X.W.)

‡Corresponding author.

  luhz@sustech.edu.cn (H.L.)

Supplemental Material for

**Anomalous magneto-thermoelectric behavior in massive Dirac materials**

Yanan Li[1,10], Huichao Wang[2], Jingyue Wang[3,11], Chunming Wang[4,5], Yanzhao Liu[1], Jun Ge[1], Jingjing Niu[3], Wenjie Zhang[3], Pinyuan Wang[1], Ran Bi[3], Jinglei Zhang[6], Ji-Yan Dai[7], Jiaqiang Yan[8], David Mandrus[8,9], Nitin Samarth[10], Haizhou Lu[5,16‡], Xiaosong Wu[3,15†], Jian Wang[1,5,12,13,14*]

[1]*International Center for Quantum Materials, School of Physics, Peking University, Beijing 100871, China.*

[2]*School of Physics, Sun Yat-sen University, Guangzhou 510275, China.*

[3]*State Key Laboratory for Artificial Microstructure and Mesoscopic Physics, Peking University, Beijing 100871, China.*

[4]*Department of Physics, Shanghai Normal University, Shanghai 200234, China.*

[5]*Institute for Quantum Science and Engineering and Department of Physics, Southern University of Science and Technology, Shenzhen 518055, China.*

[6]*High Magnetic Field Laboratory, Chinese Academy of Sciences, Hefei 230031, Anhui, China.*

[7]*Department of Applied Physics, The Hong Kong Polytechnic University, Hung Hom, Kowloon, Hong Kong, China.*

[8]*Materials Science and Technology Division, Oak Ridge National Laboratory, Oak Ridge, Tennessee 37831, USA.*

[9]*Department of Materials Science and Engineering, University of Tennessee, Knoxville, Tennessee 37996, USA.*

[10]*Department of Physics, The Pennsylvania State University, University Park, PA 16802, USA.*

[11]*Center for Nanochemistry, Beijing Science and Engineering Center for Nanocarbons, Beijing National Laboratory for Molecular Sciences, College of Chemistry and Molecular Engineering, Peking University, Beijing 100871, China.*

[12]*Collaborative Innovation Center of Quantum Matter, Beijing 100871, China.*

[13]*CAS Center for Excellence in Topological Quantum Computation, University of Chinese Academy of Sciences, Beijing 100190, China.*

[14]*Beijing Academy of Quantum Information Sciences, Beijing 100193, China.*

[15]*Beijing Key Laboratory of Quantum Devices, Peking University, Beijing 100871, China.*

[16]*Shenzhen Key Laboratory of Quantum Science and Engineering, Shenzhen 518055, China.*

Y.L., H.W., Jingyue W and C.W. contributed equally to this work.

[*]Corresponding author.

 jianwangphysics@pku.edu.cn (J.W.)

†Corresponding author.

 xswu@pku.edu.cn (X.W.)

‡Corresponding author.

 luhz@sustech.edu.cn (H.L.)

**Contents**

**I. Methods**

**Devices fabrication**

ZrTe$_5$ flakes were exfoliated with the scotch tape. Then they were transferred onto SiO$_2$/Si substrates and spin-coated with photoresists. After that, standard electron beam lithography, development, and metal evaporation were carried out to generate the electrodes and on-chip heater. For devices used in thermoelectric transport measurements, Pd (6.5 nm)/Au (40 nm) was used for the heater, and Pd (6.5 nm)/Au (300 nm) was used for other electrodes.

**Transport measurements**

Most transport measurements were performed in a 14 T helium cryostat and a 9 T Quantum Design Physical Property Measurement System by using the lock-in method. The ultrahigh field measurements were conducted in a static magnetic field facility (33 T) at the Chinese High Magnetic Field Laboratory at Hefei.

**II. Supplementary experiment data**

FIG. S1. Optical microscopy image of a ZrTe$_5$ thin flake device.

FIG. S2. Thermoelectric voltage $V_{xx}(T)$ of ZrTe$_5$ thin flakes as a function of temperature.

FIG. S3. Hall data and carrier density of ZrTe$_5$ thin flake sample f1.

FIG. S4. Hall data and carrier density of ZrTe$_5$ thin flake sample f4.

FIG. S5. Hall data and carrier density of ZrTe$_5$ thin flake sample f6.

FIG. S6. Angle dependence and temperature dependence of magneto-resistance and magneto-thermopower of ZrTe$_5$ (sample f6).

FIG. S7. Angle dependent magneto-thermopower of ZrTe$_5$ near the parallel magnetic fields (sample f1).

FIG. S8. Energy-dispersive X-ray spectroscopy results of a typical ZrTe$_5$ bulk crystal.

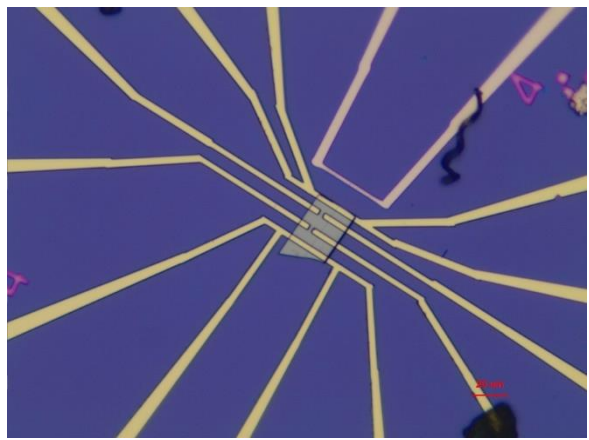

FIG. S1. Optical microscopy image of a typical ZrTe$_5$ thin flake device for the magneto-resistance and magneto-thermopower measurements.

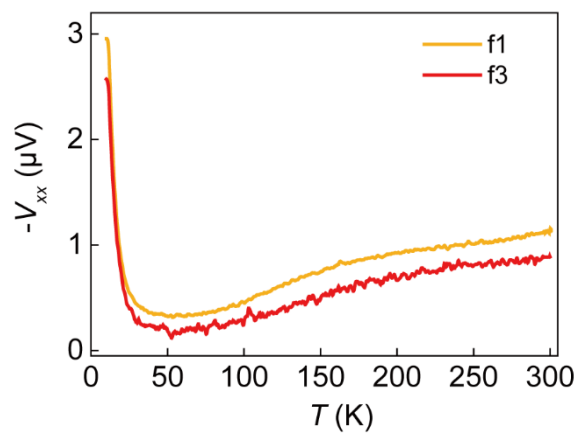

FIG. S2. Thermopower $V_{xx}(T)$ of thin flakes as a function of temperature. There is no sign change of thermopower, indicating that the dominant carrier type (hole, since $S_{xx}=-V_{xx}/\nabla T>0$) does not change in the whole measurement temperature range of 2-300 K. f1 and f3 are 250 nm and 40 nm thick, respectively.

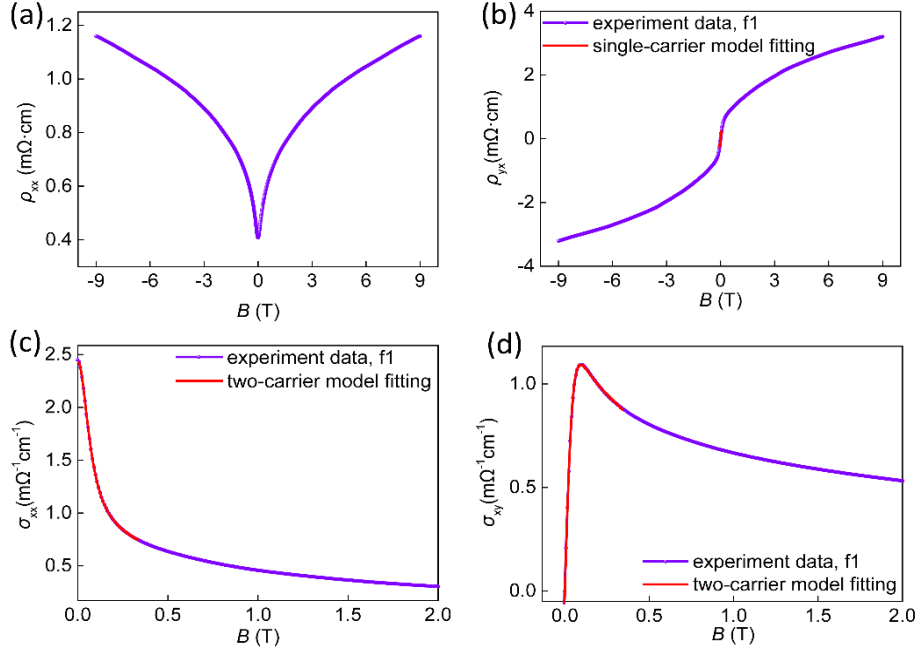

FIG. S3. Hall data and carrier density of sample f1. (a, b) Magneto-resistivity and Hall trace ($B//b$ axis) at 1.5 K. The red line in (b) is the linear fitting of the low field part (±0.05 T) of the Hall trace by considering a single-carrier model. The carrier ($p$-type) density and the mobility are estimated to be $1.3\times10^{17}$ cm$^{-3}$ and $1.2\times10^{5}$ cm$^{2}$V$^{-1}$s$^{-1}$, respectively. (c, d) Conductivity tensors calculated from resistivity tensors. The red lines are the fitting result using the two-carrier model. There are two types of hole carriers in ZrTe$_5$, and the carrier densities (mobilities) are $7.9\times10^{16}$ cm$^{-3}$ ($1.38\times10^{5}$ cm$^{2}$V$^{-1}$s$^{-1}$) and $2.46\times10^{17}$ cm$^{-3}$ ($2.6\times10^{4}$ cm$^{2}$V$^{-1}$s$^{-1}$), respectively.

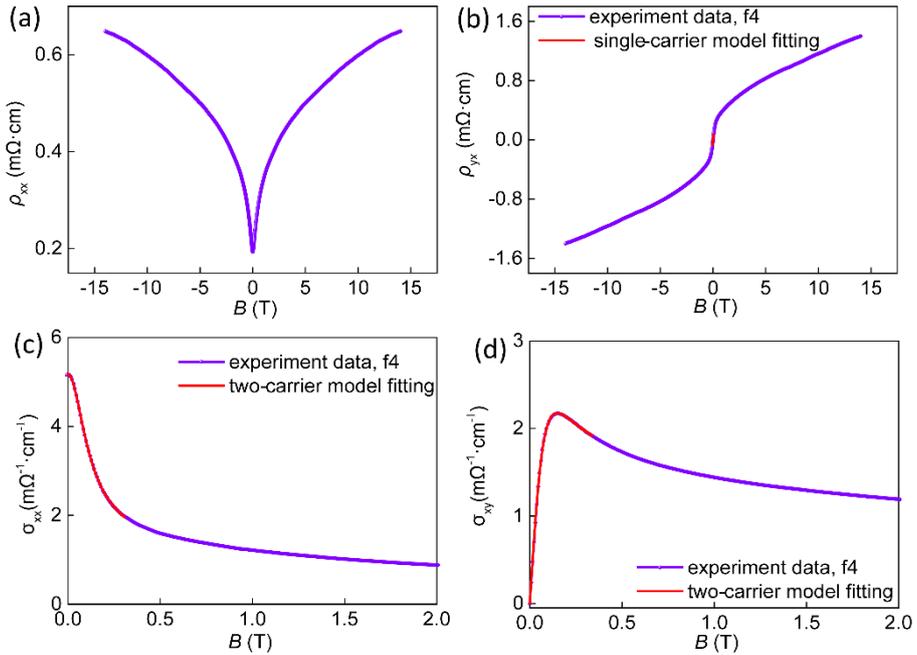

FIG. S4. Hall data and carrier density of sample f4. (a, b) Magneto-resistivity and Hall trace ($B//b$ axis) at 3 K. The red line in (b) is the linear fitting of the low field part (±0.05 T) of the Hall trace by considering a single-carrier model. The carrier (*p*-type) density and the mobility are estimated to be $5.1 \times 10^{17}$ cm$^{-3}$ and $4.31 \times 10^4$ cm$^2$V$^{-1}$s$^{-1}$, respectively. (c, d) Conductivity tensors calculated from resistivity tensors. The red lines are the fitting results using the two-carrier model. Two types of hole carriers in ZrTe$_5$ are revealed, and the carrier densities (mobility) are $2.6 \times 10^{17}$ cm$^{-3}$ ($8.46 \times 10^4$ cm$^2$V$^{-1}$s$^{-1}$) and $7.0 \times 10^{17}$ cm$^{-3}$ ($1.68 \times 10^4$ cm$^2$V$^{-1}$s$^{-1}$), respectively.

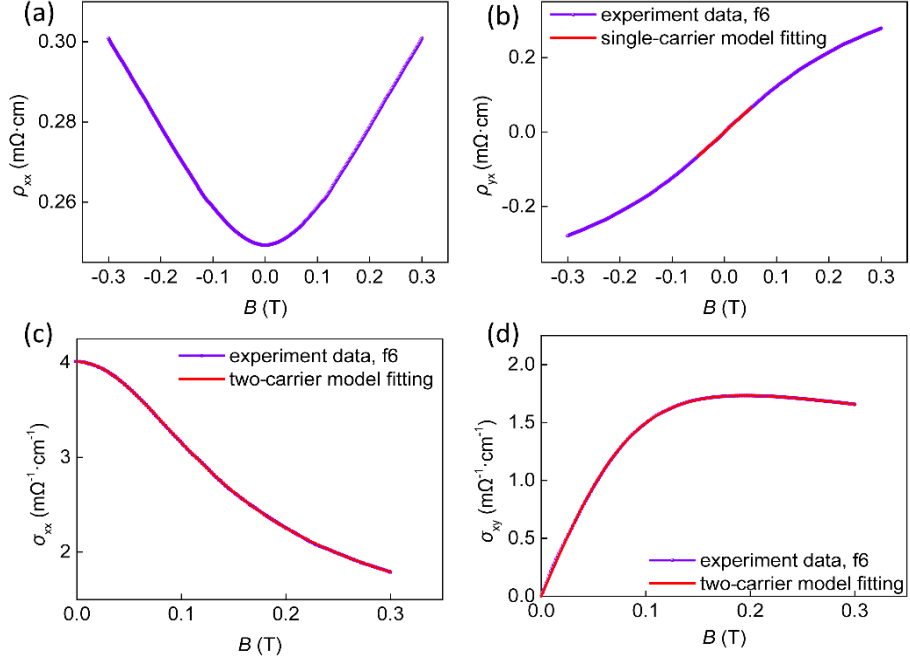

FIG. S5. Hall data and carrier density of sample f6. (a, b) Magneto-resistivity and Hall trace (*B*//*b* axis) at 3 K. The red line in (b) is the linear fitting of the low field part (±0.05 T) of the Hall trace by considering a single-carrier model. The carrier (*p*-type) density and the mobility are estimated to be $4.7 \times 10^{17}$ cm$^{-3}$ and $3.1 \times 10^4$ cm$^2$V$^{-1}$s$^{-1}$, respectively. (c, d) Conductivity tensors calculated from resistivity tensors. The red lines are the fitting results using the two-carrier model. Two types of hole carriers are revealed, and the carrier densities (mobility) are $2.35 \times 10^{17}$ cm$^{-3}$ ($6.9 \times 10^4$ cm$^2$V$^{-1}$s$^{-1}$) and $4.72 \times 10^{17}$ cm$^{-3}$ ($1.9 \times 10^4$ cm$^2$V$^{-1}$s$^{-1}$), respectively.

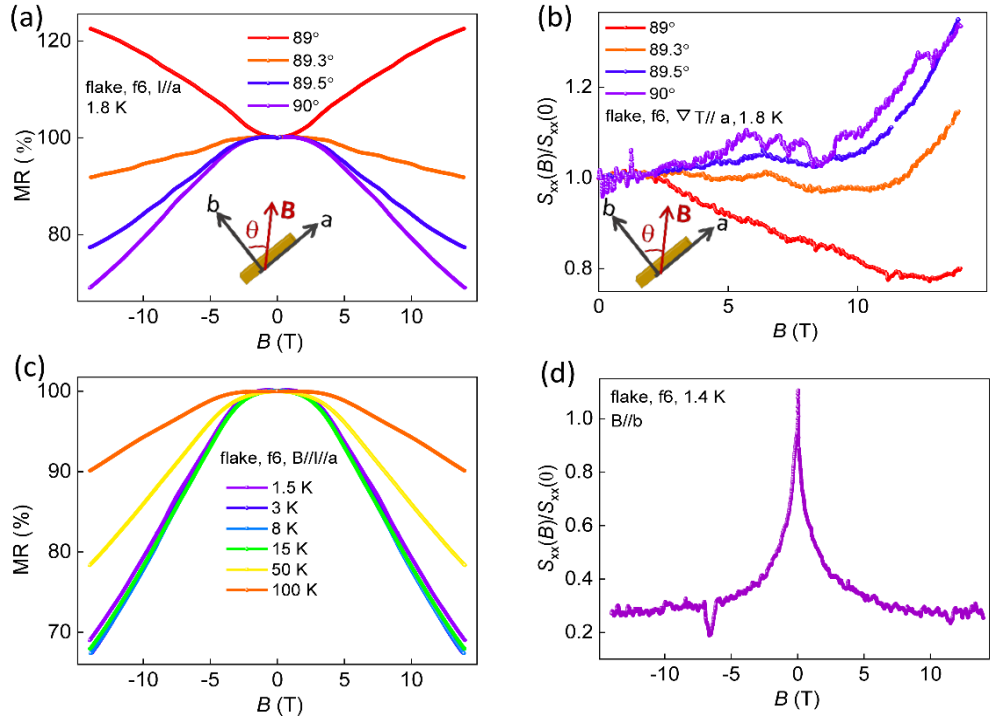

FIG. S6. Magneto-resistance and magneto-thermopower of sample f6. (a) Angular dependence of magnetoresistance in *ab* plane. When *B* and *I* are both along *a* axis, there is obvious negative MR behavior. The negative MR disappears when the external field tilts away from *a* axis. (b) Angular dependence of the magneto-thermopower in *ab* plane. Positive MTP shows up when $B//\nabla T//a$ axis. (c) Temperature dependence of negative MR when $B//a$ axis. (d) Negative MTP when $B//b$ axis.

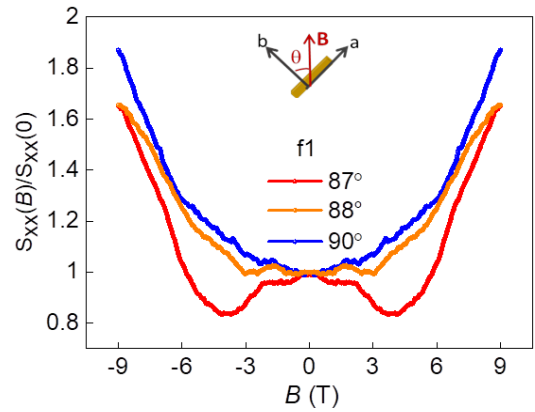

FIG. S7. Angle dependent magneto-thermopower of ZrTe$_5$ near the parallel magnetic fields (sample f1). The MTP is positive when the temperature gradient and the magnetic field are parallel ($\theta=90°$).

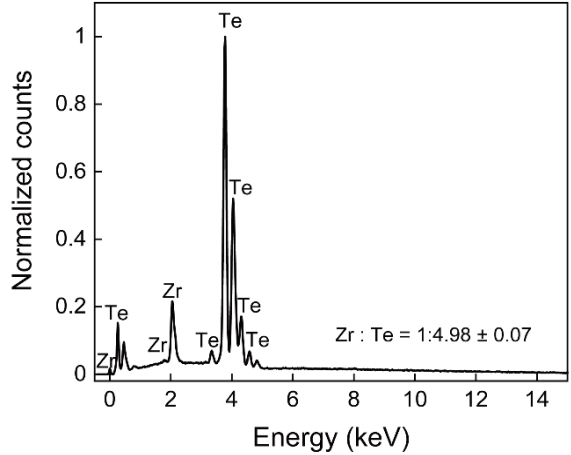

FIG. S8. Energy-dispersive X-ray spectroscopy results of a typical ZrTe$_5$ bulk crystal. The chemical composition of ZrTe$_5$ crystals is Zr: Te ~ 1: 4.98.

**III. Discussion on magneto-thermopower under parallel fields**

The flat trace at low fields for f4 (Fig. 3(c) in main text) may be due to the out-of-plane contribution from a tiny misalignment angle in the measurement because a sharp negative MTP is observed under perpendicular fields below 5 T. The normalized MTP near the parallel magnetic field of a different sample f1 with more detailed angle-dependence measurements are shown in Fig. S7. We can see that the MTP is overall positive at $\theta=90°$. When the orientation of the magnetic field is 3° away from the in-plane direction ($\theta=87°$), negative MTP is observed at low fields while the MTP remains positive field dependence at high fields. This demonstrates that slight misalignment in the measurements of parallel field configuration significantly influences the low-field MTP result but barely change the high-field behavior.

**IV.  Theoretical calculations of the transport behavior**

For the case where the magnetic field is along the $b$ axis and the electric field or temperature gradient is along the $a$ axis, we adopt a low-energy $k \cdot p$ Hamiltonian of ZrTe$_5$ [1]

$$H = vk_x\tau_x \otimes \sigma_z + vk_y\tau_y \otimes \sigma_0 + vk_z\tau_x \otimes \sigma_x + m\tau_z + b_z\sigma_z. \tag{S1}$$

Here the $x$, $y$, $z$-axes correspond to the crystal $a$, $c$, $b$-axes, respectively. $\tau = (\tau_x, \tau_y, \tau_z)$ and $\sigma = (\sigma_x, \sigma_y, \sigma_z)$ represent the Pauli matrices in the orbital and spin spaces. $v$ is the Fermi velocity, $m$ describes the mass, and $b_z = g\mu_B B/2$ is the Zeeman term induced by the external magnetic field $B$ with $g$ being the effective $g$ factor and $\mu_B$ being the Bohr magneton. This model has been widely used to discuss various experimental observations of ZrTe$_5$ bulk samples and thick flakes [1,2]. When a uniform magnetic field is applied along the $z$ direction $\boldsymbol{B} = B\hat{z}$, we take the Landau gauge $\boldsymbol{A} = -By\hat{x}$ with $\boldsymbol{A}$ the vector potential. Under the Peierls substitution, the wave vector in the above Hamiltonian is replaced by $\boldsymbol{k} \to \boldsymbol{k} - e\boldsymbol{A}/\hbar$. The eigenenergies and eigenstates of the system for $\nu = 0$ are

$$E_{0s} = s\sqrt{v^2 k_z^2 + (b_z - m)^2}, \tag{S2}$$

$$\psi_{0s} = e^{i(k_x x + k_z z)} \begin{bmatrix} 0 \\ \sin\frac{\alpha}{2}\phi_0 \\ \cos\frac{\alpha}{2}\phi_0 \\ 0 \end{bmatrix}, \tag{S3}$$

with $\tan\alpha = vk_z/(b_z - m)$. For $\nu \geq 1$, the eigeneneries are

$$E_{\nu s\lambda} = s\sqrt{\left(\sqrt{\nu\omega^2 + m^2} + s\lambda b_z\right)^2 + v^2 k_z^2}. \tag{S4}$$

The eigenstates are $\psi_{\nu s\lambda} = e^{i(k_x x + k_z z)}\left[c_{1k_z}^{\nu s\lambda}\phi_{\nu-1}, c_{2k_z}^{\nu s\lambda}\phi_\nu, c_{3k_z}^{\nu s\lambda}\phi_\nu, c_{4k_z}^{\nu s\lambda}\phi_{\nu-1}\right]^T$. Here the indices $s, \lambda = \pm 1$, the coefficients $c_{1k_z}^{\nu s\lambda}, c_{2k_z}^{\nu s\lambda}, c_{3k_z}^{\nu s\lambda}, c_{4k_z}^{\nu s\lambda}$ can be obtained from the secular equation and $\phi_\nu$ are the usual harmonic oscillator eigenstates $\phi_\nu(k_x, y) = (\sqrt{\pi} 2^\nu \nu! \ell_B)^{-1/2} e^{-[(y-y_0)^2/2\ell_B^2]} \mathcal{H}_\nu[(y-y_0)/\ell_B]$ with $y_0 = k_x \ell_B^2$, $\ell_B = \sqrt{\hbar/eB}$, and $\mathcal{H}_\nu$ being the Hermite polynomials.

In the following calculation we concentrate on the quantum limit where all the electrons occupy the lowest Landau band $E_{0s}$. In this condition, the Fermi energy $E_F$ is determined by the fixed electron density $n$ as

$$E_F = \sqrt{4\pi^4 v^2 n^2 \ell_B^4 + (b_z - m)^2}, \tag{S5}$$

and the Fermi wave vector $k_F = 2\pi^2 n \ell_B^2$.

The Seebeck coefficient is written as $S_{xx} = -E_x/|\nabla T| = \rho_{xx}\alpha_{xx} + \rho_{yx}\alpha_{xy}$. Here $\rho_{ij}$ and $\alpha_{ij}$ are the resistivities and Peltier conductivities with $i, j = (x, y)$. The thermoelectric conductivities $\alpha_{ij}$ can be related to the electric conductivities $\sigma_{ij}$ via the Mott relation. Then the Seebeck coefficient is obtained as

$$S_{xx} = \frac{\pi^2 k_B^2 T}{3e} \frac{1}{\sigma_{xx}^2 + \sigma_{xy}^2}\left(\sigma_{xx}\frac{\partial \sigma_{xx}}{\partial E_F} + \sigma_{xy}\frac{\partial \sigma_{xy}}{\partial E_F}\right) = \rho_{xx}\alpha_{xx} + \rho_{yx}\alpha_{xy}. \tag{S6}$$

In the quantum limit, the longitudinal conductivity for the positive Fermi energy is [3]

$$\sigma_{xx} = \frac{\hbar e^2}{2\pi L_z}\frac{1}{12\pi \ell_B^2}\int_{-\infty}^{\infty} d\varepsilon \left[-\frac{\partial n_F(\varepsilon)}{\partial \varepsilon}\right]\sum_{s,\lambda,k_z} \text{Re}\left[\left(v_{0,1s\lambda}^x\right)^2 G_{1s\lambda}^A(\varepsilon) G_0^R(\varepsilon)\right], \tag{S7}$$

with $n_F(\varepsilon)$ being the Fermi distribution function, the Green's functions $G_0^R(\varepsilon) = \left(\varepsilon - E_{0+} + i\frac{\hbar}{2\tau_0}\right)^{-1}$, $G_{1s\lambda}^A(\varepsilon) = \left(\varepsilon - E_{1s\lambda} - i\frac{\hbar}{2\tau_{1s\lambda}}\right)^{-1}$, and the velocity element $v_{0,\nu s\lambda}^x = \frac{v}{\hbar}\left(c_1^{\nu s\lambda}\cos\frac{\alpha}{2} - c_4^{\nu s\lambda}\sin\frac{\alpha}{2}\right)\delta_{\nu 1}$.

The longitudinal conductivity is approximated as

$$\sigma_{xx} \simeq \frac{\hbar e^2}{2\pi} \frac{1}{2\ell_B^2} \int_{-\infty}^{\infty} d\varepsilon \left[-\frac{\partial n_F(\varepsilon)}{\partial \varepsilon}\right] \sum_{s,\lambda,k_z} \left[(v_{0,1s\lambda}^x)^2 \frac{\frac{\hbar}{2\tau_{1s\lambda}}}{(\varepsilon - E_{1s\lambda})^2 + \left(\frac{\hbar}{2\tau_{1s\lambda}}\right)^2} \delta(\varepsilon - E_{0+}^{k_z})\right]. \tag{S8}$$

The relaxation time $\tau_{1s\lambda}$ due to the electron-impurity scattering can be calculated by using the Born approximation [3,4] as

$$\frac{\hbar}{\tau_{1s\lambda}} = \pi n_i \ell_B^2 \sum_{k_z'} \left(F_{k_z,k_z'}^{1s\lambda,0}\right)^2 \sum_q |u(q)|^2 q_\perp^2 e^{-\ell_B^2 q_\perp^2/2} \delta_{q_z, k_z - k_z'} \delta\left(\varepsilon - E_{0+}^{k_z'}\right). \tag{S9}$$

Here $n_i$ is the impurity density, $u(q)$ is the Fourier transformation of the scattering potential, $q_\perp^2 = q_x^2 + q_y^2$, and the form factor

$$F_{k_z,k_z'}^{1s\lambda,0} = c_{2k_z}^{1s\lambda} \sin\frac{\alpha_{k_z'}}{2} + c_{3k_z}^{1s\lambda} \cos\frac{\alpha_{k_z'}}{2}. \tag{S10}$$

In order to obtain the Seebeck coefficient, we also need the Hall conductivity, which is given by [3]

$$\sigma_{yx} = \frac{k_F}{\pi} \frac{e^2}{h} = \frac{ne}{B}. \tag{S11}$$

Its derivative is $\frac{\partial \sigma_{yx}}{\partial E_F} = \frac{e^2}{h} \frac{1}{\pi v} \frac{E_F}{\sqrt{E_F^2 - (b_z - m)^2}}$.

At zero temperature $-\partial n_F(\varepsilon)/\partial \varepsilon \to \delta(\varepsilon - E_F)$, then we only need the relaxation time at the Fermi energy $\tau_{1s\lambda}^{\pm k_F}$. For the random Gaussian scattering potential [4]

$$U(r) = \sum_i \frac{u_0}{(d\sqrt{2\pi})^3} e^{-|r - R_i|^2/2d^2}, \tag{S12}$$

where $u_0$ measures the scattering strength of a randomly distributed impurity at $R_i$, and $d$ is a parameter that determines the range of the scattering potential. The Fourier transform is

$$u(q) = u_0 e^{-q^2 d^2/2}. \tag{S13}$$

The other type of scattering is the screened Coulomb potential,

$$U(r) = \sum_i \frac{e^2}{4\pi \varepsilon_0 \varepsilon_r |r - R_i|} e^{-\kappa |r - R_i|}. \tag{S14}$$

Its Fourier transform is

$$u(q) = \frac{e^2}{\varepsilon_0 \varepsilon_r (q^2 + \kappa^2)}. \tag{S15}$$

Here $\varepsilon_r$ is the relative dielectric constant and $\kappa$ is the inverse of the screening length, which can be calculated in the standard random phase approximation [5-7]

$$\kappa^2 = -\frac{e^2}{\varepsilon_0 \varepsilon_r} \frac{1}{2\pi l_B^2} \frac{1}{\beta} \sum_m \int_{-\infty}^{+\infty} \frac{dk_z}{2\pi} \frac{1}{(i\omega_m + E_F - E_{0+})^2}$$

with $\beta = 1/k_B T$ and $\omega_m = (2m+1)\pi/\beta$ as the Matsubara frequencies of fermions. After summing over the Matsubara frequencies, we have

$$\frac{1}{\beta} \sum_m \frac{1}{(i\omega_m + E_F - E_{0+})^2} = \frac{\partial n_F(E_{0+})}{\partial E_{0+}}.$$

Substituting this into the expression of $\kappa^2$, we have at zero temperature

$$\kappa^2 = \frac{e^2}{\varepsilon_0 \varepsilon_r} \frac{1}{2\pi^2 \ell_B^2} \frac{E_F}{v^2 k_F}. \tag{S16}$$

In general, the polarizability function should be momentum- and temperature-dependent just like the one in graphene [8]. Here we consider the long wavelength limit at zero temperature.

In Fig. S9, we plot the numerically-calculated magneto-resistance and magneto-thermopower for various dielectric constants or acting ranges of the impurity potential. The MR is always positive for both potentials, which agrees with the understandings of magneto-resistance behavior [9]. The MTP in Figs. S9 (a) and (c) decreases with increasing fields for both long-range screened Coulomb potential ($1/\kappa > l_B$, $\varepsilon_r = 10 - 20$, Fig. S9 (b)) and Gaussian potential ($d > l_B$, $d$=15 - 20 nm), which is consistent with our experimental observations. However, the MTP increases with the increasing magnetic field for short-range screened Coulomb potential ($1/\kappa < l_B$, $\varepsilon_r = 0.5 - 1.5$) and short-range Gaussian potential ($d < l_B$, $d$=2, 5 nm).

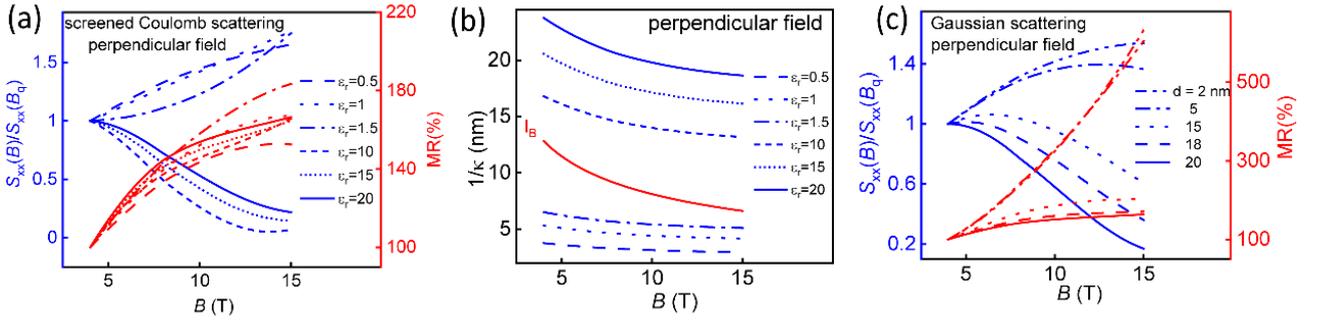

FIG. S9. (a) Calculated MTP and MR under perpendicular fields in the presence of the screened Coulomb with various dielectric constants. (b) The screening length for different dielectric constants in (a) under perpendicular fields. $\varepsilon_r = 0.5, 1, 1.5$ corresponds to short-range screened Coulomb potential while $\varepsilon_r = 10, 15, 20$ corresponds to long-range potential. The red curve represents the magnetic length $l_B$. (c) Calculated MTP and MR under perpendicular fields in the presence of the Gaussian potentials with various acting ranges. The long-range screened Coulomb potential and long-range Gaussian potential lead to positive MR and negative MTP, consistent with the experimental observations.

Now we consider the parallel situation where both the current and magnetic fields are along the $a$ axis. The uniform magnetic field takes the form $\boldsymbol{B} = B\hat{x}$ and thus the vector potential can take $\boldsymbol{A} = -Bz\hat{y}$. Therefore, the Hamiltonian is

$$H = vk_x\tau_x \otimes \sigma_z + v\left(k_y + \frac{z}{\ell_B^2}\right)\tau_y \otimes \sigma_0 + vk_z\tau_x \otimes \sigma_x + m\tau_z + b_x\sigma_x. \tag{S17}$$

To obtain the Landau bands of the above Hamiltonian, one need a unitary transformation

$$H' = (\tau_0 \otimes U^\dagger)H(\tau_0 \otimes U) = vk_x\tau_x \otimes \sigma_x + v\left(k_y + \frac{tz}{l_B^2}\right)\tau_y \otimes \sigma_0 + vk_z\tau_x \otimes \sigma_z + m\tau_z + b_x\sigma_z, \tag{S18}$$

where

$$U = \frac{\sqrt{2}}{2}\begin{pmatrix} 1 & 1 \\ 1 & -1 \end{pmatrix}. \tag{S19}$$

Then the eigenenergies and eigenstates of the system for $v = 0$ are

$$E_{0s} = s\sqrt{v^2 k_x^2 + (b_x + m)^2}, \tag{S20}$$

$$\psi_{0s} = e^{i(k_x x + k_y y)}\begin{pmatrix} i\sin\frac{\alpha}{2}\phi_0 \\ 0 \\ 0 \\ i\cos\frac{\alpha}{2}\phi_0 \end{pmatrix}, \tag{S21}$$

with $\tan\alpha = -vk_x/(b_x + m)$.

The longitudinal conductivity in the quantum limit can be calculated using the Kubo formula

$$\sigma_{xx} = \frac{\hbar e^2}{2\pi}\frac{1}{2\pi\ell_B^2}\int_{-\infty}^{\infty} d\varepsilon\left[-\frac{\partial n_F(\varepsilon)}{\partial\varepsilon}\right]\sum_{k_x}(v_0^x G_0^A v_0^x G_0^R), \tag{S22}$$

with the velocity element $v_0^x = \frac{v}{\hbar}\sin\alpha$. The advanced Green's function is $G_0^A(\varepsilon) = \left(\varepsilon - E_{0+} - i\frac{\hbar}{2\tau_0}\right)^{-1}$. We use the weak-scattering approximation,

$$G_0^R G_0^A \simeq \frac{2\pi\tau}{\hbar}\delta(\varepsilon - E_{0+}). \tag{S23}$$

It is known that the vertex correction of the velocity is equivalent to correct the relaxation time into the transport time. The transport time at the Fermi surface is calculated as

$$\frac{\hbar}{\tau_{k_F}^{tr}} = 2\pi\sum_{k_x',k_y'}\left\langle\left|U_{k_F,k_y;k_{x'},k_{y'}}^{0,0}\right|^2\right\rangle_{imp}\left(1 - \frac{v_{0,k_x'}^x}{v_{0,k_F}^x}\right)\delta\left(E_F - E_{0+}^{k_x'}\right), \tag{S24}$$

with $\left\langle\left|U_{k_F,k_y;k_{x'},k_{y'}}^{0,0}\right|^2\right\rangle_{imp}$ denoting the scattering matrix element and $v_{0,k_x'}^x = \frac{v}{\hbar}\frac{vk_x'}{E_{0+}^{k_x'}}$. Finally, the conductivity is written as

$$\sigma_{xx} = \frac{e^2}{h}\frac{1}{2\pi\ell_B^2}\frac{v^2 k_F}{E_F}\left(\frac{\tau_{k_F}^{tr}}{\hbar} + \frac{\tau_{-k_F}^{tr}}{\hbar}\right). \tag{S25}$$

Here the transport time is

$$\frac{\hbar}{\tau_{k_F}^{tr}} = \frac{2E_F}{v^2 k_F}n_i\cos^2\left[\frac{1}{2}(\alpha_{k_F} - \alpha_{-k_F})\right]\sum_{q_y,q_z}|u(2k_F,q_y,q_z)|^2 e^{-q_\perp^2\ell_B^2/2},$$

$$\frac{\hbar}{\tau^{tr}_{-k_F}} = \frac{2E_F}{v^2 k_F} n_i \cos^2\left[\frac{1}{2}(\alpha_{k_F} - \alpha_{-k_F})\right] \sum_{q_y, q_z} |u(-2k_F, q_y, q_z)|^2 e^{-q_\perp^2 \ell_B^2/2}.$$

(S26)

The resistivity and the thermopower from the Mott formula are

$$\rho_{xx} = \frac{1}{\sigma_{xx}},$$
$$S_{xx} = \frac{\pi^2 k_B^2 T}{3e} \frac{1}{\sigma_{xx}} \frac{\partial \sigma_{xx}}{\partial E_F}.$$

(S27)

For the random Gaussian potential, the thermopower is

$$S_{xx} = \frac{\pi^2 k_B^2 T}{3e} \frac{2E_F}{E_F^2 - (b_x + m)^2}\left\{1 + \frac{4d^2}{v^2}[E_F^2 - (b_x + m)^2]\right\},$$

(S28)

and the resistivity is

$$\rho_{xx} = \frac{h}{e^2} \frac{n_i u_0^2}{v^2} \frac{\ell_B^2}{2d^2 + \ell_B^2} \frac{(b_x + m)^2}{E_F^2 - (b_x + m)^2} e^{-4[E_F^2 - (b_x + m)^2]d^2/v^2}.$$

(S29)

For fixed $E_F$, the $B$-dependence of $S_{xx}$ is only in $b_x$. However, in the quantum limit, only the $E_{0+}$ band is occupied, the fixed one should be the electron density $n$, instead of $E_F$. For the fixed density, the resistivity is

$$\rho_{xx} = \frac{h}{e^2} \frac{n_i u_0^2}{v^2} \frac{(b_x + m)^2}{4\pi^4 v^2 n^2 \ell_B^2 (2d^2 + \ell_B^2)} e^{-16\pi^4 n^2 \ell_B^4 d^2}.$$

(S30)

Here, $n_i$, $u_0$, $v$, $b_x$, $m$, $n$ are impurity density, scattering strength, Fermi velocity, Zeeman energy, mass and carrier density, respectively. As the magnetic field is increased, $b_x$ increases, $l_B$ decreases, both the fraction and the $e$-exponential functions increase, and then the resistivity always increases with increasing field for Gaussian scattering with any acting range. This is opposite to the experimental observations. Further, for the screened Coulomb impurity potential, the thermopower and resistivity are

$$S_{xx} = \frac{\pi^2 k_B^2 T}{3e}\left\{\frac{2E_F}{E_F^2 - (b_x + m)^2} - \frac{\zeta + e^\zeta \zeta^2 \text{Ei}(-\zeta) - 1}{\zeta + e^\zeta \zeta^2 \text{Ei}(-\zeta)}\left(\frac{4\ell_B^2}{v^2} E_F - \frac{e^2}{4\pi^2 \epsilon v}\frac{(b_x + m)^2}{[E_F^2 - (b_x + m)^2]^{3/2}}\right)\right\}$$

(S31)

and

$$\rho_{xx} = \frac{h}{e^2} \frac{n_i e^4}{4\epsilon^2 v^2} \frac{(b_x + m)^2}{E_F^2 - (b_x + m)^2}\left[\frac{1}{\zeta} + e^\zeta \text{Ei}(-\zeta)\right]\ell_B^4.$$

(S32)

Here $\text{Ei}(x)$ is the exponential integral and $\zeta$ is

$$\zeta = \frac{2\ell_B^2}{v^2}[E_F^2 - (b_x + m)^2] + \frac{e^2}{4\pi^2 \epsilon v}\frac{E_F}{\sqrt{E_F^2-(b_x+m)^2}}. \tag{S33}$$

The screened Coulomb scattering could give the negative MR and the positive MTP. We have checked and found that it is valid even for the screening length smaller than the magnetic length as shown in Fig. S10. Therefore, the specific momentum-dependence of the screened Coulomb potential leads to the observed negative MR and positive MTP.

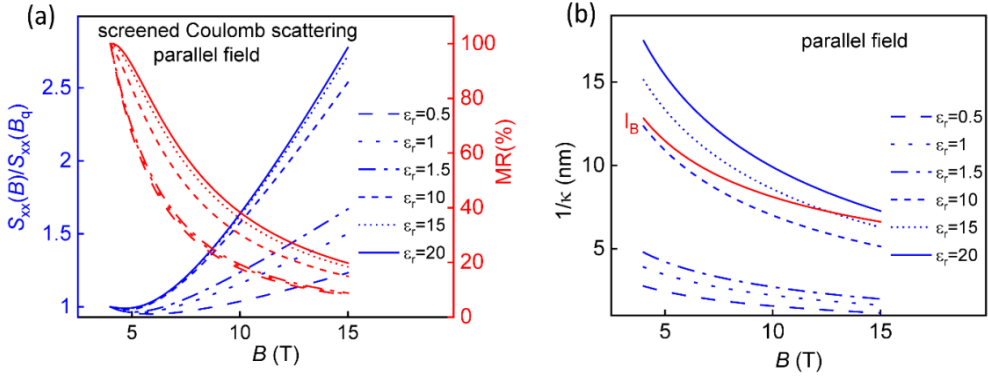

FIG. S10. (a) The MR and MTP for screened Coulomb scattering with various dielectric constants in the parallel fields. MR shows negative field dependence while MTP shows positive field dependence for both long-range and short-range screened Coulomb potentials ($\varepsilon_r$ = 0.5–20). (b) The screening length under parallel fields. $\varepsilon_r$ = 0.5, 1, 1.5 corresponds to short-range potentials while $\varepsilon_r$ = 20 corresponds to long-range one under parallel fields.

By combining the theoretical calculations and the experimental observations, one can judge that the dominant scattering in our samples is the long-ranged screened Coulomb potential. The parameters used in the numerical calculations are, the perpendicular field Fermi velocity ($v_{F\perp}$) = 0.3 eVnm [10], parallel Fermi velocity ($v_{F\parallel}$) = 0.09 eVnm [10], mass $m$ = 0.009 eV [11], g-factor $g$ = 22.5 [1]. The plot in Fig. 4 of the main text corresponds to $\varepsilon_r$ = 15 and $d$ = 20 nm.

Here we also show the calculations with a two-carrier Drude model for perpendicular fields. It is found that both long-range screened Coulomb scattering and Gaussian scattering lead to positive MR and negative MTP under the perpendicular fields, which is consistent with the above calculation results. The longitudinal and Hall conductivities are written as

$$\sigma_{xx} = n_1 e\mu_1 \frac{1}{1+(\mu_1 B)^2} + n_2 e\mu_2 \frac{1}{1+(\mu_2 B)^2},$$

$$\sigma_{yx} = n_1 e\mu_1 \frac{\mu_1 B}{1+(\mu_1 B)^2} + n_2 e\mu_2 \frac{\mu_2 B}{1+(\mu_2 B)^2}. \tag{S34}$$

Here $n_{1,2}$ and $\mu_{1,2}$ are the densities and mobilities of two carriers. Then the resistivity and the Seebeck

coefficient are

$$\rho_{xx} = \frac{\sigma_{xx}}{\sigma_{xx}^2 + \sigma_{yx}^2},$$

$$S_{xx} = \frac{\pi^2 k_B^2 T}{3e} \frac{\sigma_{xx}\sigma'_{xx} + \sigma_{yx}\sigma'_{yx}}{\sigma_{xx}^2 + \sigma_{yx}^2}. \tag{35}$$

Here the prime means the derivative versus the Fermi energy. The conductivity of one type of carrier in the absence of magnetic field is ($a = 1,2$)

$$\sigma_{a,xx} = n_a e \mu_a = e \sum_{\mathbf{k}} v_{a,x}^2 \tau^{tr}(E_{a,\mathbf{k}}) \left(-\frac{\partial n_F(E_{a,\mathbf{k}})}{\partial E_{a,\mathbf{k}}}\right), \tag{S36}$$

with the energy $E_{a,\mathbf{k}} = \epsilon_a - \sqrt{v_a^2 k^2 + m}$ and $v_{a,x} = \hbar^{-1} \partial E_{a,\mathbf{k}}/\partial k_x$ being the group velocity. We consider two types of hole carriers, and they have different Fermi velocity $v_a$ and top of the band $\epsilon_a$. At zero temperature, the mobility reduces to

$$\mu_a = \sigma_{a,xx}/(n_a e) = e \frac{v_a^2}{\hbar^2} \frac{\tau_{a,F}^{tr}}{E_F}. \tag{S37}$$

The transport time of the $a$-th hole at Fermi energy $\tau_{a,F}^{tr}$ is

$$\frac{\hbar}{\tau_{a,F}^{tr}} = \frac{n_i}{4\pi} \frac{E_F k_{a,F}}{v_a^2} \int_0^\pi d\theta \left|u(q_{a,F}(\theta))\right|^2 \sin\theta (1-\cos\theta)(1+\cos\theta), \tag{S38}$$

with $k_{a,F}$ being the Fermi wave vector of the $a$-th hole and $q_{a,F}(\theta) = 2k_{a,F}\sin\frac{\theta}{2}$. For the screened Coulomb scattering, the transport time is calculated as

$$\frac{\hbar}{\tau_{a,F}^{tr}} = \frac{n_i e^4}{8\pi\epsilon^2} \frac{E_F}{v_a^2 k_{a,F}^3} \left[\left(\frac{\kappa_a^2}{2k_{a,F}^2} + 1\right) \ln \frac{\frac{\kappa_a^2}{4k_{a,F}^2} + 1}{\frac{\kappa_a^2}{4k_{a,F}^2}} - 2\right]. \tag{S39}$$

The inverse of the screening length is determined by

$$\kappa_a^2 = -\frac{e^2}{\epsilon} \sum_{\mathbf{k}} \frac{\partial n_F(E_{a,\mathbf{k}})}{\partial E_{a,\mathbf{k}}} = \frac{e^2}{\epsilon} \frac{1}{2\pi^2} \frac{E_F k_{a,F}}{v_a^2}. \tag{S40}$$

And for the gaussian potential, it is

$$\frac{\hbar}{\tau_{a,F}^{tr}} = n_i u_0^2 \frac{E_F k_{a,F}}{\pi v_a^2} \frac{1}{16 k_{a,F}^6 d^6} \left[2k_{a,F}^2 d^2 - 1 + (1 + 2k_{a,F}^2 d^2) e^{-4k_{a,F}^2 d^2}\right]. \tag{S41}$$

The Seebeck coefficient becomes

$$S_{xx} = \frac{\pi^2 k_B^2 T}{3e}(f_1 + f_2). \tag{S42}$$

Here

$$f_1 = \frac{n_1^{4/3}\mu_1^2(1+(\mu_2 B)^2) + \left(n_1 n_2^{1/3} + n_1^{1/3} n_2\right)\mu_1\mu_2(1+\mu_1\mu_2 B^2) + n_2^{4/3}\mu_2^2(1+(\mu_1 B)^2)}{n_1^2\mu_1^2(1+(\mu_2 B)^2) + 2n_1\mu_1 n_2\mu_2(1+\mu_1\mu_2 B^2) + n_2^2\mu_2^2(1+(\mu_1 B)^2)},$$

$$f_2 = \rho_{xx}\sigma_{xx}\left\{\frac{n_1\mu'_1(1+(\mu_2 B)^2)[n_1\mu_1(1+(\mu_2 B)^2) + n_2\mu_2(1-(\mu_1 B)^2 + 2\mu_1\mu_2 B^2)]}{[n_1\mu_1(1+(\mu_2 B)^2) + n_2\mu_2(1+(\mu_1 B)^2)]^2} + \right.$$

$$\frac{n_2\mu_2'(1+(\mu_1 B)^2)[n_2\mu_2(1+(\mu_1 B)^2)+n_1\mu_1(1-(\mu_2 B)^2+2\mu_1\mu_2 B^2)]}{[n_1\mu_1(1+(\mu_2 B)^2)+n_2\mu_2(1+(\mu_1 B)^2)]^2}\bigg\}. \tag{S43}$$

Fig. S11 shows the calculated MR and MTP under perpendicular fields with two-carrier Drude model in the presence of the long-range screened Coulomb scattering and long-range Gaussian scattering. The negative MTP and the positive MR are obtained.

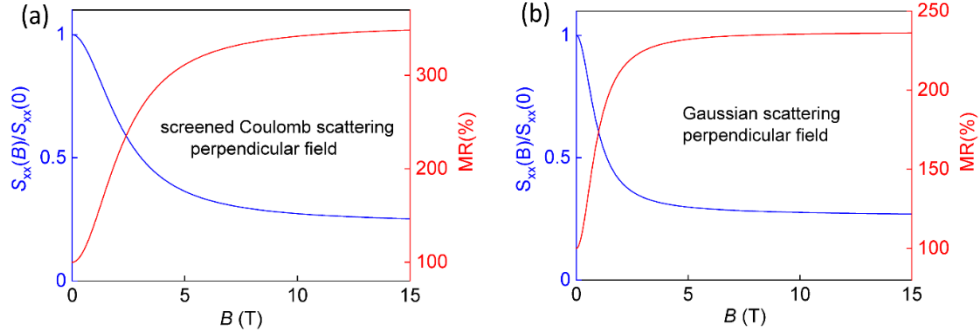

FIG. S11. The calculated MR and MTP with a two-carrier Drude model in the presence of long-range screened Coulomb scattering (a) and long-range Gaussian scattering (b) under perpendicular fields. The calculation parameters are, the Fermi velocities $v_1 = 0.3\,\text{eVnm}$ and $v_2 = 0.1\,\text{eVnm}$, the mass $m = 0.009\,\text{eV}$, the dielectric constant $\varepsilon_r = 15$, and the acting range of the Gaussian potential $d = 20\,\text{nm}$.